\documentclass[page-classic]{epl2}

\usepackage{amssymb}
\usepackage{amsmath}
\usepackage{amsfonts}
\usepackage{graphicx}
\usepackage{bbm}
\usepackage{color}
\title{Nonperturbative signatures in pair production for general elliptic polarization fields}

\shorttitle{Nonperturbative signatures in pair production for general elliptic polarization fields} %Insert here a short version of the title if it exceeds 70 characters

\author{Z. L. Li\inst{1} \and D. Lu\inst{1} \and B. S. Xie\inst{1,2}\footnote{Corresponding author: bsxie@bnu.edu.cn} \and B. F. Shen\inst{3} \and L. B. Fu\inst{4} \and J. Liu\inst{4}}
\shortauthor{Z. L. Li \etal}

\institute{
  \inst{1} Key Laboratory of Beam Technology and Materials Modification of the Ministry of Education, College of Nuclear Science and Technology, Beijing Normal University, Beijing 100875, China\\
  \inst{2} Beijing Radiation Center, Beijing 100875, China\\
  \inst{3} Shanghai Institute of Optics and Fine Mechanics, Chinese Academy of Sciences, Shanghai 201800, China\\
  \inst{4} National Laboratory of Science and Technology on Computational Physics,
Institute of Applied Physics and Computational Mathematics, Beijing 100088, China}

\pacs{12.20.Ds}{Specific calculations}
\pacs{11.15.Tk}{Other nonperturbative techniques}
\pacs{32.80.-t}{Photon interactions with atoms}
\pacs{42.50.Xa}{Optical tests of quantum theory}

\abstract{The momentum signatures in nonperturbative multiphoton pair production for general elliptic polarization electric fields are investigated by employing the real-time Dirac-Heisenberg-Wigner formalism. For a linearly polarized electric field we find that the positions of the nodes in momenta spectra of created pairs depend only on the electric field frequency. The polarization of external fields could not only change the node structures or even make the nodes disappear but also change the thresholds of pair production. The momentum signatures associated to the node positions in which the even-number-photon pair creation process is forbid could be used to distinguish the orbital angular momentum of created pairs on the momenta spectra. These distinguishable momentum signatures could be relevant for providing the output information of created particles and  also the input information of ultrashort laser pulses.}

\begin{document}

\maketitle

\section{Introduction}

Electron-positron (EP) pair creation in strong electric fields \cite{Sauter,Heisenberg,Schwinger} is a fundamentally important exploration to probe the quantum vacuum. As a famous prediction of quantum electrodynamics, experimental observation of this nonperturbative tunneling pair production is still absent due to the fact that present available laser fields are far below the Schwinger critical electric field $E_{\mathrm{cr}}=m^2/ e \sim1.32\times10^{16}\mathrm{V/cm}$, where $m$ is the electron rest mass and $-e$ is the electron charge (we use $\hbar=c=1$). However, with the rapid development of laser technology, the stronger fields will be achieved in near future through some laser facilities \cite{ELI,XFEL,Ringwald}. This improves greatly the hopes to realize the observation of EP pair production from vacuum \cite{Piazza1, Bell,Piazza2,Bulanov,Abdukerim2013,Li2014}.
For long time two different pair production mechanisms are identified by the nonperturbative pair creation process ($\gamma\ll1$) and the perturbative multiphoton pair production process ($\gamma\gg1$), where $\gamma=m\omega/eE_0$ is the well known Keldysh adiabatic parameter \cite{Keldysh}, $\omega$ and $E_0$ are the frequency and strength of external electric fields, respectively. Fortunately, the perturbative multiphoton process has been experimentally observed at the Stanford Linear Accelerator Center (SLAC) \cite{Burke}.

Compared to cases of $\gamma\ll1$ and $\gamma\gg1$, obviously the case of $\gamma\sim\mathcal{O}(1)$ is seldom studied because of the difficulty to get a simple asymptotic formulae theoretically. However, more and more evidence indicates that the pair production in this intermediate regime is an important and interesting topic because it contains the signatures of multiphoton process and nonperturbative mechanism. Many novel phenomena are discovered and deepen the understanding of pair creation mechanisms \cite{Ruf,Schutzhold,GVDunne,Kohlfurst2014}. For instance, the dynamically assisted Schwinger mechanism \cite{Schutzhold,GVDunne} by the superimposition of a strong and slowly varying electric field (nonperturbative mechanism) and a weak and rapidly changing one (multiphoton process) can dramatically enhance the pair creation rate. Recently the effective mass in the nonperturbative multiphoton pair creation regime \cite{Kohlfurst2014} has been pointed out which will not only explain the peak positions of momentum distributions but also likely be observed directly in the laboratory.

In this Letter, our studies are focused on the momentum signatures in the nonperturbative multiphoton pair creation for general elliptical polarized electric fields. By using the real-time Dirac-Heisenberg-Wigner (DHW) formalism \cite{Bialynicki,Hebenstreit2010,Hebenstreit2011}, we first revisit the momenta spectra of created pairs in a linearly polarized field and exhibit the node structures in the momenta spectra \cite{Mocken}. The interference carpets is found, which appears also in the atomic above-threshold ionization problem \cite{Korneev2012}. By studying the positions of nodes, we find that they depend only on the external field frequency rather than the Keldysh parameter. Further, when the elliptic polarized fields are considered, the node structures will change. Even the nodes will disappear for large values of polarization. Finally, we make the relation clearer about the node positions associated to the absence of even-number-photon pair creation. This study, on one hand, is valuable to deepen the understanding of both nonperturbative mechanism and perturbative multiphoton process in arbitrary polarization fields. On the other hand, it is also helpful to understand other relevant physical processes, such as cosmological pair production \cite{Parker1968}, heavy ion collisions \cite{Kharzeev2007}, and ionization of atoms and molecules \cite{Borca2001,Shafir2013}, especially the above-ionization with elliptically polarized laser pulses \cite{Bashkansky1987,Paulus1998}.

\section{Theoretical formalism}
Here we consider the field in an antinode of the standing wave formed by two counterpropagating laser pulses with appropriate polarizations. Since the spatial scales of the EP pair production are smaller than the spatial focusing scales of the laser pulse, the spatial effects are not significant and the magnetic field is absent. Therefore, we have a spatially homogeneous and time-dependent electric field. For the general polarization the field can be written as
\begin{equation}\label{eq1}
\mathbf{E}(t)=\frac{E_0}{\sqrt{1+\delta^2}}\exp\Big(-\frac{t^2}{2\tau^2}\Big)[                                              \cos(\omega t+\phi),\delta\sin(\omega t+\phi), 0]^\textsf{T},
\end{equation}
where $E_0$ is the maximal field strength, $\tau$ defines the pulse duration, $\omega$ is the laser frequency, $\phi$ is the carrier phase, and $-1\leq\delta\leq1$ represents the polarization of the electric field, the factor $\sqrt{1+\delta^2}$ is used to ensure the same laser intensity for different polarization fields. For convenience, we set $\tau=100/m$, $\phi=0$ and $\delta\geq0$ throughout this paper.

Our numerical results are based on the DHW formalism which has been used to study the vacuum pair production for different electric fields, refer to the references \cite{Bialynicki,Hebenstreit2010,Hebenstreit2011}. Particularly, Ref. \cite{Hebenstreit2010} showed that the DHW formalism would be recovered to the quantum Vlasov equation (QVE) if the spatially homogeneous and time-dependent electric fields are linearly polarized. However one of the advantages of the DHW formalism is that it can also solve more complex electric fields such as (\ref{eq1}). In order to precisely obtain the momentum distribution function of created EP pairs $f$ for the spatially homogeneous electric field (\ref{eq1}), we adopt the trick used in \cite{Blinne} and the DHW formalism is reduced to
\begin{eqnarray}\label{eq2}
&\dot{f}=1/2~\dot{\mathbbm{e}}_1^\textsf{T} \mathcal{F}\mathbbm{w}_9,&  \nonumber\\
&\dot{\mathbbm{w}}_9=\mathcal{H}_9\mathbbm{w}_9+2(1-f)\mathcal{G}\dot{\mathbbm{e}}_1.&
\end{eqnarray}
Here the dot denotes a total time derivative, $\mathbbm{e}_1=(m/\Omega(\mathbf{p}), \mathbf{p}/\Omega(\mathbf{p}), \mathbf{0}, \mathbf{0})^\textsf{T}$, $\Omega(\mathbf{p})=(m^2+\mathbf{p}^2)^{1/2}$ is the total energy of electrons with the kinetic momentum ${\mathbf{p}=\mathbf{q}-e\mathbf{A}(t)}$, $\mathbf{q}$ is the canonical momentum, $\mathbf{A}(t)$ is the vector potential; $\mathbbm{w}_9$ represents an auxiliary $9$-component vector;
$\mathcal{F}=\left(
               \begin{array}{cccccc}
                 -\mathbf{p}^\textsf{T}/m ~~ \mathbf{0} \\
                 ~~~\mathbbm{1}_9 \\
               \end{array}
             \right)
$ is a $10\times9$ matrix, $\mathcal{G}=(\mathbf{0}~~\mathbbm{1}_9)$ is a $9\times10$ matrix, and
\begin{equation}
\mathcal{H}_9=\left(
                \begin{array}{ccc}
                  -e\mathbf{p}\cdot \mathbf{E}^\textsf{T}/\omega^2(\mathbf{p}) & -2\mathbf{p}\times & -2m \\
                  -2\mathbf{p}\times & \mathbf{0} & \mathbf{0} \\
                  2(m^2+\mathbf{p}\cdot \mathbf{p}^\textsf{T})/m & \mathbf{0} & \mathbf{0} \\
                \end{array}
              \right)\nonumber
\end{equation}
is a $9\times9$ matrix.
Thus, we can get the one-particle momentum distribution function $f(\mathbf{q},t)$ by solving Eq. (\ref{eq2}) with the initial conditions $f(\mathbf{q},-\infty)=\mathbbm{w}_9(\mathbf{q},-\infty)=0$.

\section{Results}
\begin{figure}[htpb]\suppressfloats
\includegraphics[width=13cm]{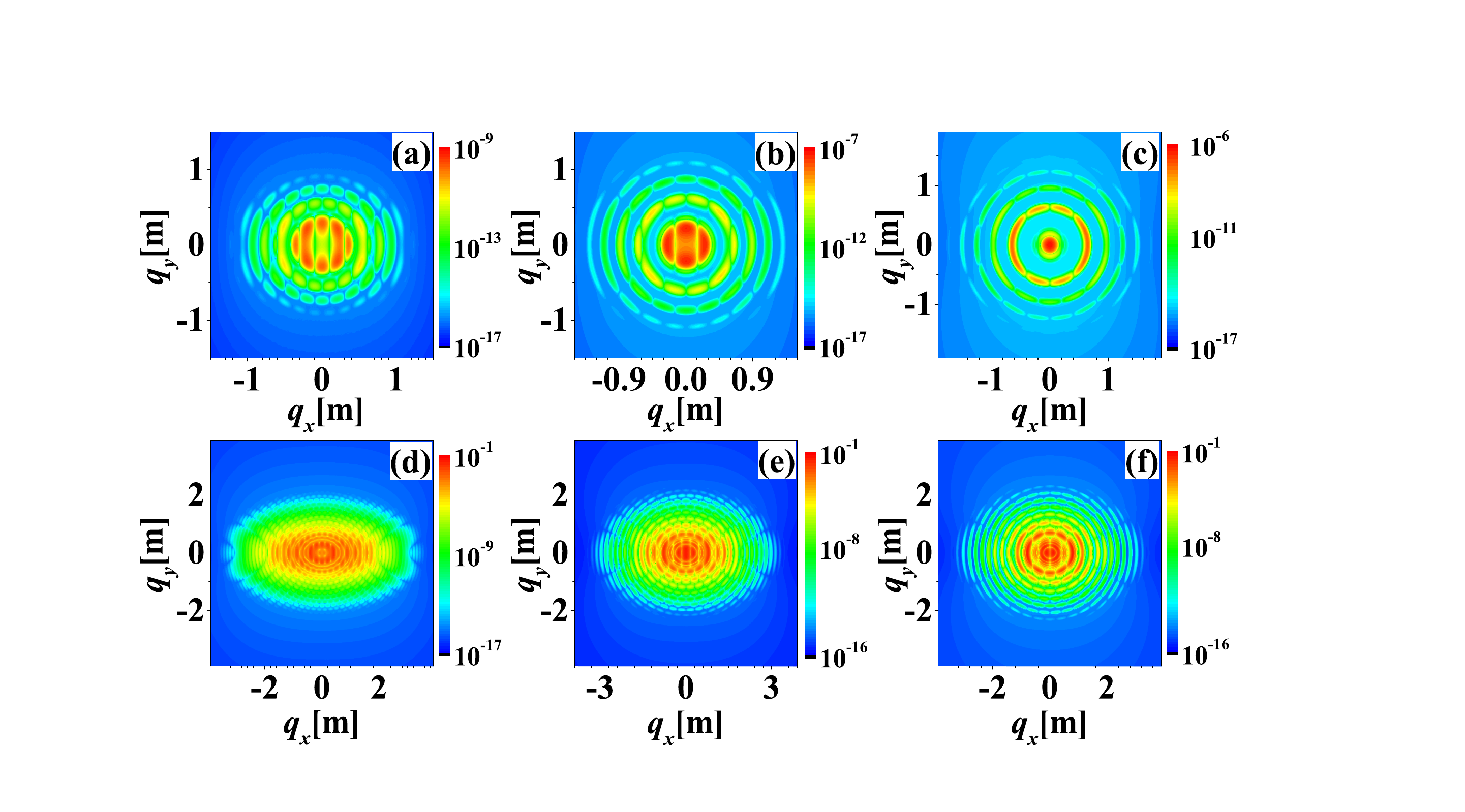}
\caption{\label{Fig1}(color online) Momenta spectra of created EP pairs in the $(q_x, q_y)$ plane for linearly polarized electric field. The upper row for $E_0=0.1E_\mathrm{cr}$ and the lower row for $E_0=0.4E_\mathrm{cr}$. The panels in each row, from left to right, are for $\omega=0.2m$, $0.3m$, and $0.4m$, respectively.}
\end{figure}

By solving Eq. (\ref{eq2}), we plot the momenta spectra of created EP pairs in Fig. \ref{Fig1} for linearly polarized electric field ($\delta=0$) with different Keldysh parameters, from (a) to (f), $\gamma=2$, $3$, $4$, $0.5$, $0.75$, and $1$.
%It is found that the ring-like structures which are caused by the multiphoton absorption become more and more pronounced as $\gamma$ increases. Furthermore, comparing the lower row with the upper one, one can see that for a larger electric field strength, the pair creation by absorbing much photon numbers is present, and the ring corresponding to a small photon-number absorption shrinks or even disappears. The reason for the latter case is due to the fact that for a fixed laser frequency, the effective mass, i.e. $m_*=m\sqrt{1+e^2E_0^2/(2m^2\omega^2)}$ \cite{Kohlfurst2014}, grows with the electric field strength, so the radius of the ring for $n$-photon process, $q=\sqrt{(n\omega/2)^2-m_*^2}$, will decrease. Particularly the ring structure will vanish completely when the field strength becomes high enough. This phenomenon is similar to the channel closing in atomic ionization \cite{Kopold2002}.
One can see node structures in the rings of momenta spectra of created particles by using the more realistic electric field (\ref{eq1}) with a Gaussian envelope. From Fig. \ref{Fig1}(f) where $\gamma=1$, we find that the nodes are located at the positions where the momentum $q_x$ is an even multiple of half the frequency for an even-number photon pair creation (cf. the $8$-photon ring with $10$ nodes), however, the nodes are located at the positions where $q_x$ is an odd multiple of half the laser frequency for an odd-number photon process (cf. the $7$-photon ring with $8$ nodes). This result can be explained via the factor $[1+(-1)^{n+2s}\cos{(4q_x/\omega\cdot\arctan{\gamma})}]$ (where $s$ is the spin of created particles, $s=0$ for bosons and $s=1/2$ for fermions, $n$ is the absorbed photon number), in the approximate analytical solution of the pair production probability for a sinusoidal field with an infinite period \cite{Popov1974,Comment0}. Furthermore, from the factor, one can see that the node positions depend on the Keldysh parameter.

\begin{figure}
\includegraphics[width=13cm]{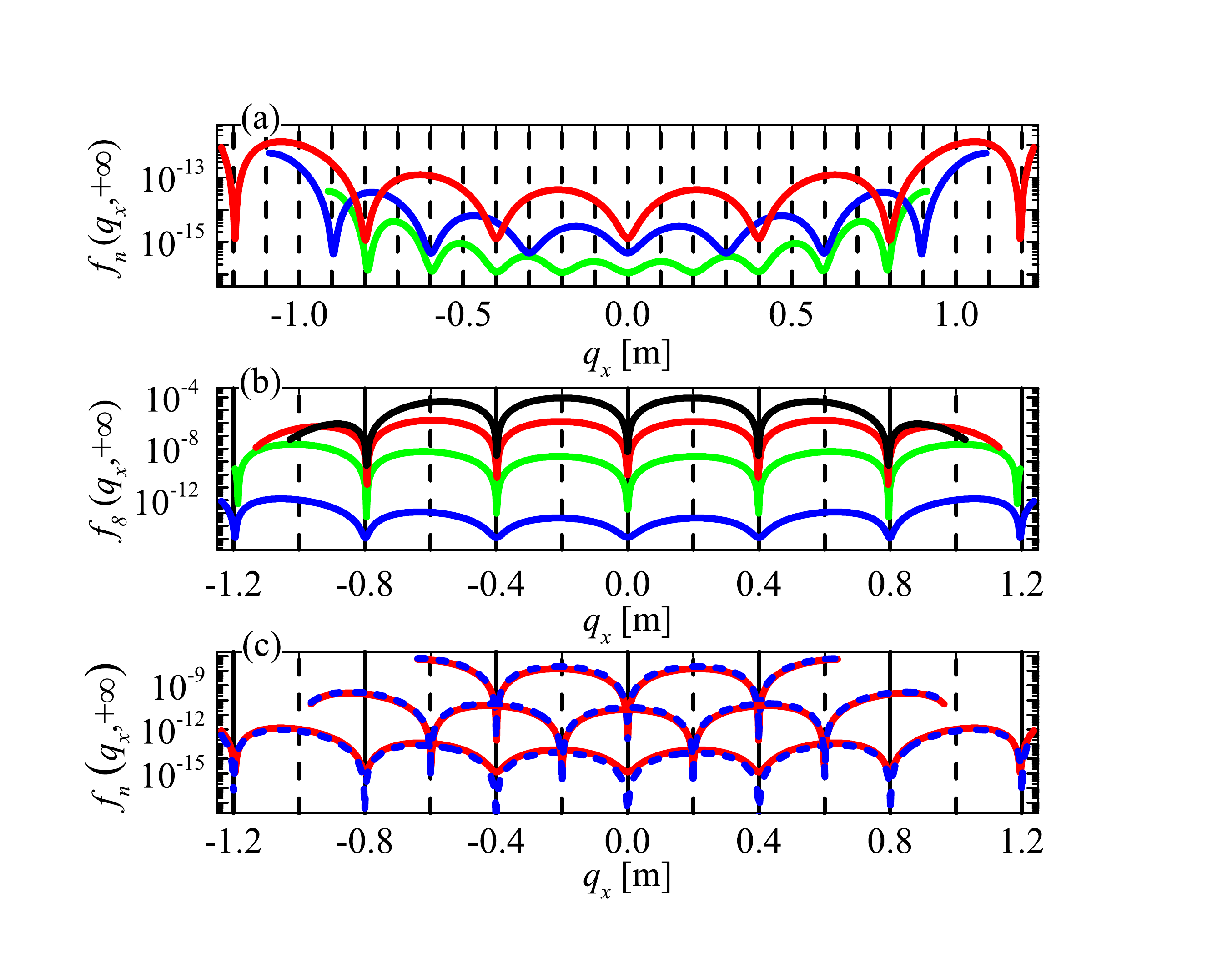}
\caption{\label{Fig2}(color online) Momentum distribution function of $n$-photon pair creation $f_n(q_x,+\infty)$ as a function of momentum $q_x$ ($q_y\geq0,~q_z=0$). (a) for $n=8$ of Fig. \ref{Fig1}(c), $10$ of Fig. \ref{Fig1}(b), and $14$ of Fig. \ref{Fig1}(a), from top to bottom. (b) corresponds to $8$-photon process with the fixed frequency $\omega=0.4m$ and $E_0/E_\mathrm{cr}=$ $0.4$, $0.3$, $0.2$, and $0.1$, from top to bottom. (c) shows the comparisons between the numerical results (solid red lines) and the approximate analytical ones (dashed blue lines) with the parameter $\mathcal{Q}=0.88$ for $n=6$, $7$, and $8$ in Fig. \ref{Fig1}(c), from top to bottom.}
\end{figure}

To carefully analyze our results shown in Fig. \ref{Fig1}, we plot the momentum distribution function of $n$-photon pair creation $f_n(q_x,+\infty)$ as a function of momentum $q_x$ ($q_y\geq0,~q_z=0$) in Fig. \ref{Fig2}. Obviously, (a) indicates that the nodes (the minimum values) of even-number photon pair creation are still located at the positions where the values of $q_x$ are integral multiples of the frequency for different $\omega$ and the fixed $E_0$. Furthermore, in (b), we find that for a given frequency the node positions will not be changed by the electric field strength. These results imply that the positions of the nodes in the momenta spectra depend \textit{only} on the field frequency rather than the values of $\gamma$. This still holds true for a sinusoidal field without Gaussian envelope. Considering this result and the effective mass signatures \cite{Kohlfurst2014} in nonperturbative multiphoton pair production, the approximate analytical expression of particle momentum distributions \cite{Popov1974} can be modified as \cite{Comment1}
\begin{eqnarray}\label{eq3}
f_n(\mathbf{q},+\infty) \sim \frac{2\omega^2}{\pi} w(\mathbf{q})[1+(-1)^{n+2s}\cos{\beta^*}]\nonumber\\
\times\delta(2\Omega_{\mathrm{rms}}(\mathbf{q})-n\omega),
\end{eqnarray}
where $w(\mathbf{q})=\exp\{-\pi E_{\mathrm{cr}}/E_0\cdot[g(\gamma)
+\mathcal{Q} b_1(\gamma)(q_y^2+q_z^2)/m^2]+ b_2(\gamma)q_x^2/m^2\}$ with $b_2(\gamma)=-\gamma b_1'(\gamma)$ and $b_1(\gamma)=g(\gamma)+\gamma g'(\gamma)/2$; the prime represents $d/d\gamma$, and $g(\gamma)=\frac{4}{\pi}\int_0^1(1-u^2)^{1/2}/[1+(\gamma u)^2]^{1/2}du$; $\mathcal{Q}$ is a parameter used to inclue the contributions of the high order terms of momentum; $\beta^*=2\pi q_x/\omega$ and $\Omega_{\mathrm{rms}}(\mathbf{q})=(\mathbf{q}^2+m_*^2)^{1/2}$ is the effective energy of electrons. The semianalytical results (dashed blue lines) are shown in Fig. \ref{Fig2} (c) and compared with the DHW results (solid red lines). We can see that these two results are agreement with each other very well \cite{Comment1}. In addition, our result is valuable to precisely measure the frequency of external fields with or without envelopes by analyzing the node positions in momentum spectra. For instance, the distance between any two adjacent nodes in a ring along the $q_x$ direction is exactly the field frequency. Our findings can also explain why the spacing between the neighbouring peaks in the center of the longitudinal momentum distribution $f(q_x,+\infty)$ is equal to the laser frequency (cf. Fig. 2 in Ref. \cite{Hebenstreit2009}). When the electric field parameters in \cite{Hebenstreit2009} are chosen as $E=0.1E_\mathrm{cr},\tau=100/m$ and $\sigma=\omega\tau=5$, one can find that $\gamma\sim1$, and the pair production is belong to the nonpertubative multiphoton regime. Therefore, the center of momenta spectra will be split according to Eq. (\ref{eq3}). A similar result can also be seen in Fig. \ref{Fig1}(d) or (e).

\begin{figure}
\includegraphics[width=13cm]{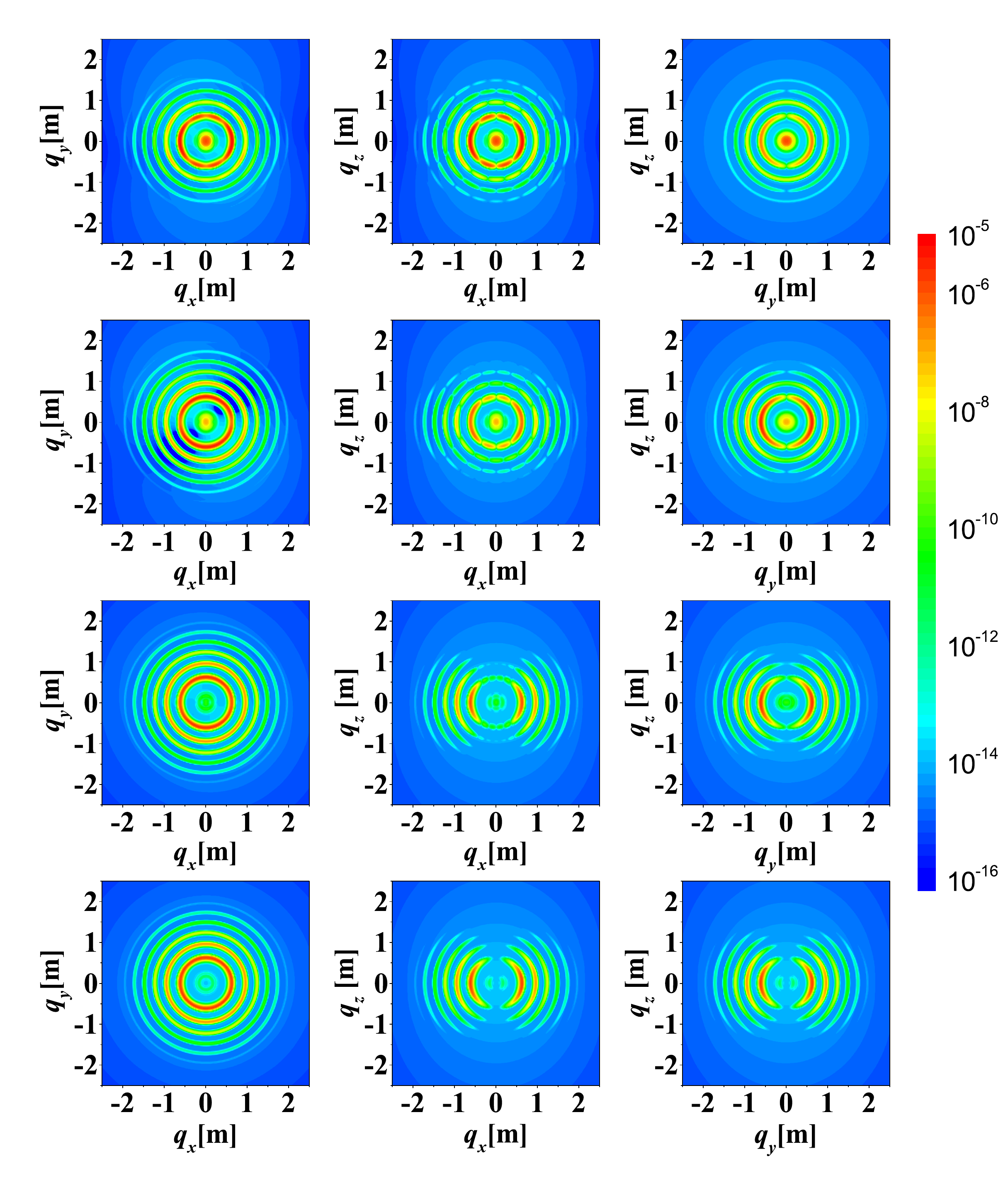}
\caption{\label{Fig3}(color online) Momenta spectra of created EP pairs for different polarizations. Each column from top to bottom corresponds to $\delta=0.1$, $0.5$, $0.9$, and $1$, respectively. Each row from left to right shows the momentum spectrum in the $(q_x, q_y)$ plane, in the $(q_x, q_z)$ plane, and in the $(q_y, q_z)$ plane, respectively. Other electric field parameters are chosen as $E_0=0.1\sqrt{2}E_\mathrm{cr}$, and $\omega=0.4m$.}
\end{figure}

For arbitrary polarized electric fields, the momenta spectra of created pairs in the $(q_x, q_y)$ plane with $q_z=0$ are shown in Fig. \ref{Fig3}. It is found that the node structures in the particle momenta spectra gradually disappear as the polarization increases, and the rings become continuous and uniform for the circular polarization $\delta=1$. This is because the rapidly rotating electric fields are more isotropic in the $(x, y)$ plane. Additionally, we find that the number of created particles in the center of momenta spectra decreases as the polarization increases, and reaches a minimum at $\delta=1$. Meanwhile it is found that the rings shrink with the polarization increasing. For example, the radius of the ring corresponding to $6$-photon pair production (the smallest ring) is about $0.628m$ for $\delta=0$, while about $0.568m$ for $\delta=1$. These indicate that a large value of polarization $\delta$ increases the thresholds of multiphoton pair creation for a fixed field frequency. Although the electric fields we used can give the same expression of the effective mass, $m_*=m\sqrt{1+e^2E_0^2/(2m^2\omega^2)}$, for different polarizations according to \cite{Kohlfurst2014}, the thresholds still grow with the increase of polarizations. This is beyond the effective mass model in \cite{Kohlfurst2014}.

In addition, we also present the momenta spectra for different polarizations in the $(q_x, q_z)$ plane with $q_y=0$ and in the $(q_y, q_z)$ plane with $q_x=0$, respectively, see Fig. \ref{Fig3}. In the $(q_x, q_z)$ plane, one can see that for a very small value of $\delta$, the node structures are pronounced and their positions can be determined approximately by Eq. (\ref{eq3}). However, when the value of $\delta$ becomes large, the field $E_y$ will be very strong, then the node structures will be disturbed by the combined effects of the field $E_x$ and $E_y$, and finally disappear for a large value of polarization. In the $(q_y, q_z)$ plane, it shows that a small value of $\delta$ corresponds to a small value of $E_y$, thus the node structures are not obvious except near $q_y=0$. With the polarization increasing, the larger rings are present, and a channel along the $z$-axis is gradually opened up near $q_y=0$.

\begin{figure}
\includegraphics[width=13cm]{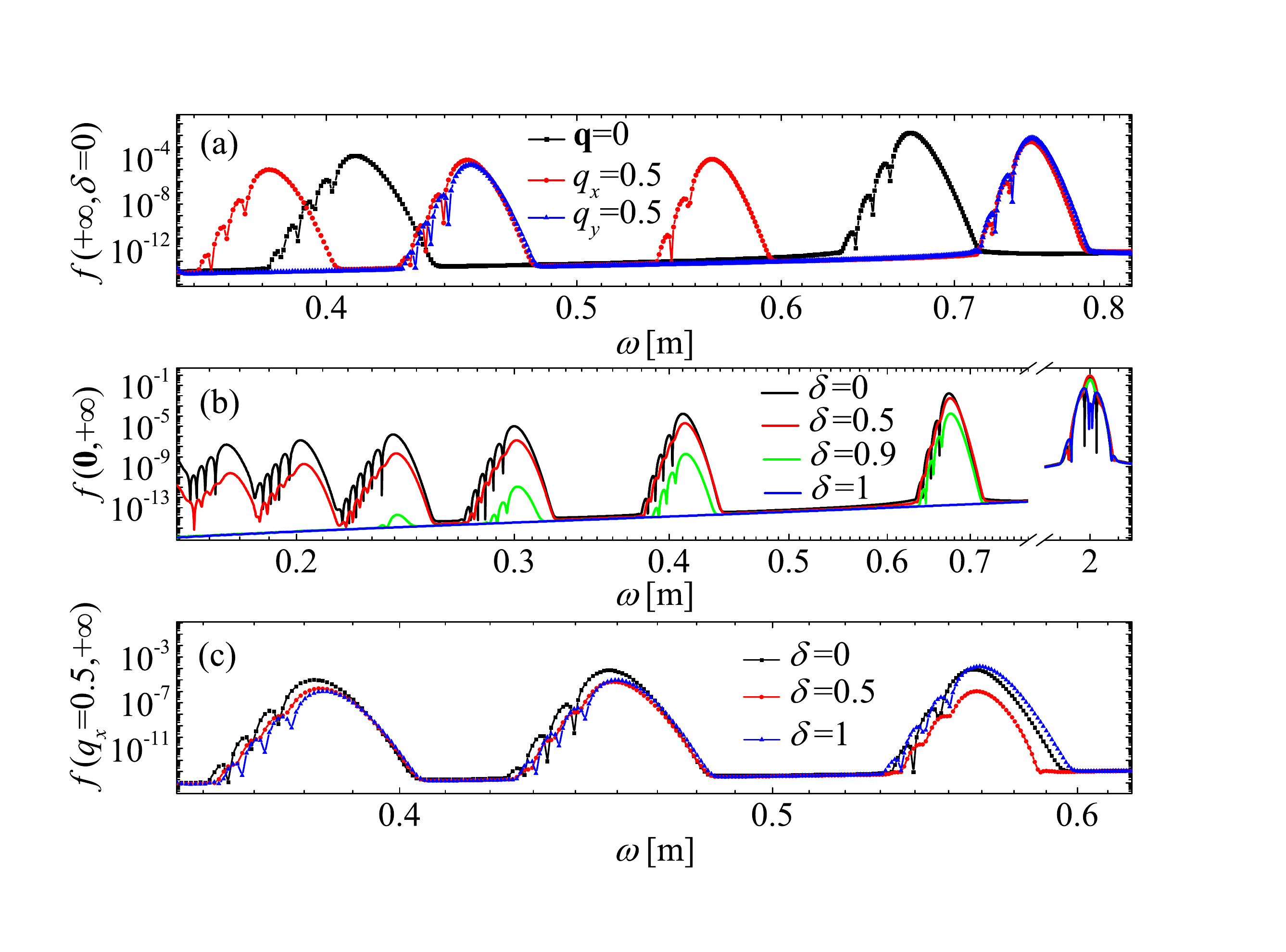}
\caption{\label{Fig4}(color online) Momentum distribution function $f(+\infty)$ as a function of electric field frequency $\omega$. (a) for linearly polarized electric field with $\mathbf{q}=0$, $q_x=0.5m$, and $q_y=0.5m$. Other unmentioned momenta are zero. (b) for different polarizations with zero momentum. From top to bottom, $\delta=0$, $0.5$, $0.9$, and $1$, respectively. (c) for $\delta=0$, $0.5$, and $1$, with $q_x=0.5m,q_y=q_z=0$.}
\end{figure}

To study the relation between the node structures in momenta spectra and the absence of even-number photon pair production, the distribution function $f(+\infty)$ as a function of the laser frequency is plotted in Fig. \ref{Fig4}. By analyzing Fig. \ref{Fig4}(a), we find that for a linearly polarized field $\delta=0$, once the momentum component along the external field is zero, namely $q_x=0$, the even-number photon pair creation will not occur. A similar result can also be seen for $\delta\neq0$, i.e., the even-number photon pair creation vanishes at $q_x=q_y=0$, see Fig. \ref{Fig4}(b) and (c). The former result can be explained quantitatively employing Eq. (\ref{eq3}). When $q_x=0$, the expression becomes $f\propto[1 + (-1)^{n+1}]$ for EP pair production. This shows that the momentum distribution function has a minimum for even-number photon pair creation, i.e., the even-number photon process vanishes. Another result can also be interpreted qualitatively applying Eq. (\ref{eq3}). However, since there are two electric fields $E_x$ and $E_y$ in the $(x,y)$ plane, the momentum $q_x$ in (\ref{eq3}) should be reasonably replaced by $(q_x^2+q_y^2)^{1/2}$. Then one can find that once $q_x=q_y=0$, the even-number photon process will not appear. The above results indicate that the absence of even-number photon pair creation at $q_x=0$ for $\delta=0$ and $q_x=q_y=0$ for $\delta\neq0$ is just one example of the node structures whose positions can be well explained by Eq. (\ref{eq3}). A more physical interpretation of the absence of even-number photon pair creation in the case of zero momentum is according to the C-parity selection rule \cite{Mocken}. Based on this explanation, we can deepen the understanding of the node structures shown in Fig. \ref{Fig1} by assuming that the orbital angular momentum (OAM) of created pairs $l$ is related to $q_x/\omega$ as $l\sim k q_x/\omega\sim k q_x/(2\pi /\lambda)\sim k\bar{\lambda} q_x$ ($k$ is an integer and $\bar{\lambda}$ is the reduced laser wave length). For instance, when $l$ is odd, the C-parity of EP pairs is $1$, then the even-number photon process having the C-parity $(-1)^{even}=1$ is possible, while the odd-number photon process having the C-parity $(-1)^{odd}=-1$ is prohibited. If the assumption turns out to be true, one can obtain the OAM for the first time by analysing the particle momentum spectra.
%Moreover, from Fig. \ref{Fig4}, one can see some oscillations on the left side of $n$-photon peaks. This is caused by the combined effect between Schwinger mechanism and multiphoton pair production. It is known that a field frequency which is less than the resonant frequency is unable to overcome the threshold of pair creation $2m_*$. However, for some frequencies a negative-energy electron in the Dirac sea can be stimulated to an intermediate state and become a positive-energy electron by tunneling process. Therefore, there will present some peaks of the distribution function for these frequencies. Another noticeable phenomenon presented in
Moreover, Fig. \ref{Fig4}(b) also presents that for the electric field with a large value of $\delta$, the resonance pair production corresponding to a large photon number can not occur. Particularly, when $\delta=1$, only one-photon pair creation is present. It seems that a rapidly rotating electric field will restrain the large photon-number pair production for zero momentum case.

\section{Conclusions}
We have investigated momentum signatures in nonperturbative multiphoton pair production for general elliptic polarized electric fields by using the real-time DHW formalism. It is found that the node positions in momenta spectra depend only on the field frequency for linear polarization and will be changed by other polarizations. Furthermore, the thresholds of multiphoton pair creation grow with the increase of polarization under the same laser intensity. The momentum signatures not only possibly provide us with the OAM of created particles but also can present some signatures of ultrashort laser pulses, such as the frequency. We believe that these phenomena are expected to be observed more easily in atom ionization for a general elliptical polarized laser pulses, since the basic physical picture in the pair production is similar to that in atomic ionization problems but obviously the laser field is available for the latter.

\acknowledgments
This work was supported by the National Natural Science Foundation of China (NSFC) under Grant Nos. 11475026, 11175023 and 11335013, and also supported partially by the Open Fund of National Laboratory of Science and Technology on Computational Physics at IAPCM and the Fundamental Research Funds for the Central Universities (FRFCU).


\begin{thebibliography}{10}
\bibitem{Comment0}
Note that in order to correctly predict the node positions in the case of $\gamma=1$, the number $4$ in the factor should be replaced by $8$, refer to \cite{Mocken}.
\bibitem{Comment1}
Note that due to the effect of a Gaussian envelope and the finite period, the delta function in Eq. (\ref{eq3}) is too ideal to exactly discrible the ring widths in momentum specta. Therefore, the delta function is just used to estimate the positions of the maximum values of $f_n(\mathbf{q})$. To compare with the numerical result, the strength of delta function is choosen as $\mathcal{P}=923$. Obviously $\mathcal{P}$ also includes the contributions of higher order terms of momentum.
\end{thebibliography}
\end{document}